\documentclass[twocolumn]{aastex62}
\turnoffedit

\usepackage{float}
\usepackage{wrapfig}
\usepackage{graphicx}
\usepackage{csvsimple}
\usepackage{units}
\usepackage[caption=false]{subfig}

\newcommand{\fearth}{$F_\oplus$}

\definecolor{dg}{HTML}{228B22}

\graphicspath{{./}}

\received{August 15, 2018}
\revised{September 20, 2018}
\accepted{October 01, 2018}

\submitjournal{ApJS}

\shorttitle{Revised Exoplanet Parameters Using Gaia DR2}
\shortauthors{Johns et al.}

\begin{document}

\title{Revised Exoplanet Radii and Habitability Using Gaia Data Release 2}
\author{Daniel Johns}
\affiliation{Department of Physical Sciences, Kutztown University, Kutztown, PA 19530, USA}

\author{Connor Marti}
\affiliation{Department of Astronomy, Williams College, Williamstown, MA 01267, USA}

\author{Madison Huff}
\affiliation{Department of Physics, Westminster College, New Wilmington, PA 16172, USA}

\author{Jacob McCann}
\affiliation{Department of Physical Sciences, Kutztown University, Kutztown, PA 19530, USA}

\author{Robert A. Wittenmyer}
\affiliation{University of Southern Queensland, Centre for Astrophysics, Toowoomba, Queensland 4350, Australia}

\author{Jonathan Horner}
\affiliation{University of Southern Queensland, Centre for Astrophysics, Toowoomba, Queensland 4350, Australia}

\author{Duncan J. Wright}
\affiliation{University of Southern Queensland, Centre for Astrophysics, Toowoomba, Queensland 4350, Australia}

\begin{abstract}
Accurate stellar properties are crucial for determining exoplanet characteristics. Gaia DR2 presents revised distances, luminosities, and radii for 1.6 billion stars. Here, we report the calculation of revised radii and densities for 320 \edit1{non-Kepler} exoplanets using this data and present updated calculations of the incident flux received by 690 known exoplanets. This allows the likelihood that those planets orbit in the habitable zone of their host stars to be reassessed. As a result of this analysis, three planets can be added to the catalogue of potentially habitable worlds: HIP~67537~b, HD~148156~b, and HD~106720~b. In addition, the changed parameterisation of BD~+49~898 means that its planet, BD~+49~898~b, now receives an incident flux that places it outside the optimistic habitable zone region, as defined by \citep{Kopparapu2013,Kopparapu2014}. We find that use of the new \textit{Gaia} data results in a mean increase in calculated exoplanet radius of 3.76\%. Previously, CoRoT-3 b had been reported as having the highest density of all known exoplanets. Here, we use updated information to revise the calculated density of CoRoT-3~b from 26.4$g\:cm^{-3}$ to 16.1$\pm3.98g\:cm^{-3}$. We also report the densest exoplanet in our dataset, KELT-1~b, with a density of 22.1$^{+5.62}_{-9.16}g\:cm^{-3}$.  Overall, our results highlight the importance of ensuring the the parameterisation of known exoplanets be revisited whenever significant improvements are made to the precision of the stellar parameters upon which they are based.

\end{abstract}

\section{Introduction} \label{sec:intro}
Over the three decades since the discovery of the first planets around other stars marked the dawn of the Exoplanet Era \citep{GammaCeph,Latham,psr1257,51peg}, we have come to realise that planets are ubiquitous, and that the variety of planetary properties and system architectures is far greater than we ever imagined \citep[e.g.][]{kepler36,20782,kelt9,trappist1}.

Aside from the unexpected worlds found orbiting the pulsar PSR1257 +12 \citep{psr1257}, the first exoplanets found were all behemoths, comparable in size to Jupiter \citep[e.g.][]{lathamsworld,47uma,70vir}. Those first planets included the first surprise of the Exoplanet Era -- the 'Hot Jupiters' -- planets the mass of Jupiter moving on orbits with periods measured in hours, or just a few days \citep[e.g.][]{51peg,HotJ2,HotJupiters}.  

In the decades since those first discoveries, the surprises have kept coming. A great diversity of alien worlds has been revealed. Some planets move in tightly packed planetary systems \citep[e.g.][]{Pack1,Pack2,Trapp}, whilst others move on extremely elongated orbits \citep[e.g.][]{80606,ecc2,ecc3}. 

In the early years of the Exoplanet Era, the predominant method used to find exoplanets was the radial velocity technique \citep[e.g.][]{RV1, RV2, RV3}. Using that technique, it is possible to constrain the orbit and mass of newly discovered planets, but with radial velocity observations alone, we can learn nothing more about the planet's physical nature.

The advent of large scale transit surveys, such as the Wide Angle Search for Planets \citep[WASP;][]{wasp1,HotJ2,wasp3}, the Hungarian Automated Telescope Network \citep[HATnet][]{HAT1,HAT2,HAT3} and the Kilodegree Extremely Little Telescope \citep[KELT][]{kelt1,KELT2,KELT3} offer a solution to this problem. 

If a planet is known to transit its host star, then its diameter can be determined. Simply - a larger planet will block more light than a smaller one, resulting in a deeper, more pronounced transit. Spectroscopic observations carried out during an exoplanet's transit can yield information on the atmospheric scale and even composition of that planet's atmosphere \citep[e.g.][]{atmos1,atmos2,atmos3}. 

In addition to such observations, measuring the radial velocity variations of the transiting planet's host star will yield its mass. As a result, it is possible to more fully characterise the planet - calculating its bulk density. Such observations have revealed an incredible breadth of potential planetary densities, ranging from planets less dense than cotton candy \citep[e.g.][]{Candy} to others denser than Osmium \citep[e.g.][]{kelt1,dense1}.

In the coming years, the focus of planet search programs will shift from primarily finding large planets to the search for potentially habitable, Earth-like worlds \citep[e.g.][]{ExoEarth1,ExoEarth2,ExoEarth3}. NASA's Transiting Exoplanet Survey Satellite \citep{TESS} should yield hundreds of such worlds over the coming years, and the race will be on to determine which, if any, could be potentially suitable as targets for the search for life beyond the Solar system \citep[e.g.][]{HabReview,Hab2,Hab3}. 

A key component of that characterisation effort will be attempts to quantify the incident flux a given planet will receive from its host star, to see whether it falls in the putative 'habitable zone', and could therefore have a reasonable likelihood of hosting liquid water upon its surface \citep[though a variety of other factors will also have to be taken into account - see e.g.][]{HabReview,FoF3,FoF4}. 

To address this, \citet{Kopparapu2013,Kopparapu2014} propose the concepts of the 'optimistic' and 'conservative' habitable zones, whose inner boundaries are defined as the limits where a planet would resemble a younger Venus, potentially harboring liquid water, and where a planet would lose its water oceans entirely to evaporation, respectively. Using these boundaries, researchers are already proposing potential systems that are worthy of attention with the next generation of radial velocity facilities as potential hosts of habitable worlds \citep[e.g.][]{Matt1,Matt2}

The characterisation of the density and potential habitability of newly discovered exoplanets depends strongly on the precision with which the host star can be characterised. The measurement of the planet's diameter, for example, relies on an accurate value for the star's size, whilst the investigation of the planet's potential habitability (in terms of the insolation received) depends critically on an accurate measurement of the star's luminosity \citep[e.g.][]{Kopparapu2014}.

For this reason, when new and improved data become available allowing planet host stars to be better characterised, it is vital that the catalog of known exoplanets be revisted, in order to ensure that the parameters available to researchers are as accurate and up-to-date as possible.

The {\it Gaia} spacecraft is currently undertaking a five-year program of observations, through which it will obtain exquisitely precise measurements of several billion stars \citep[e.g.][]{Gaia,GaiaDR1}. {\it Gaia's} observations will yield precise distances, luminosities and effective temperatures for its target stars that represent a vast improvement on the data previously available. 

The second {\it Gaia} data release was made available on April 25, 2018, and contains data for a total of almost two billion sources\footnote{Details of the {\it Gaia} Data Release 2 can be found at https://www.cosmos.esa.int/web/gaia/dr2}, of which over 1.3 billion include parallax determinations, allowing the distances to those stars to be accurately determined.

In this work, we take advantage of the recent {\it Gaia} Data Release 2 \citep{BigGaia,MoreGaia,GaiaPhotometry,GaiaSpectroscopy,GaiaParallax} to update the calculated sizes, densities, and incident fluxes for a large sample of known exoplanets. Our paper is structured as follows: in Section ~\ref{Methods}, we describe our methodology, before presenting and discussing our results in Section ~\ref{results}. Finally, in Section ~\ref{conclusion}, we draw our conclusions, and highlight the important role that surveys such as {\it Gaia} will play in the development of exoplanetary science in the coming years.

\section{Methodology} \label{Methods}

\subsection{\textit{Gaia} DR2 pipeline}
The \textit{Gaia} collaboration has made use of the Astrophysical Parameters Inference System (Apsis) to provide estimates of stellar effective temperature, luminosity, and radius for 77 million stars \citep{GaiaApsis}. Using the GSP-Phot software \citep{GaiaApsis} were able to infer stellar effective temperatures using different G-band colors. Stellar luminosities were inferred using stellar G-band magnitude, effective stellar temperature, temperature dependent bolometric correction, and the Solar bolometric magnitude, defined as $M_{bol\odot} = 4.74$. Stellar radii were then easily inferred using the Stefan-Boltzmann Law. \cite{GaiaApsis} found that estimates of extinction, $A_G$, were not accurate on a star-to-star basis and chose to set $A_G=0$. Their radii estimates show little disagreement ($<$5\%) when compared to interferometry and astroseismology literature radii for several well-studied stars. \edit1{Uncertainty in $T_{eff}$ could potentially produce large errors in stellar radius, so we choose to report the median uncertainty in the inferred effective temperature of our sample to be +1.67 \% and -1.49 \%.} Although the \textit{Gaia} DR2 luminosities tend to be underestimated, \cite{GaiaApsis} provides a means of accounting for an accurate measure of extinction or a change in bolometric correction through a set of simple exponential scaling laws. These estimated stellar luminosities and radii were used in this work. 

\subsection{Cross Matching \& Filtering}
First, we cross-matched the details of 3,735 confirmed exoplanets\footnote{Details of the confirmed exoplanets were taken from the NASA Exoplanet Archive, at https://exoplanetarchive.ipac.caltech.edu/, on 22 June.}
with an ADQL Query of the \textit{Gaia} DR2 archive. 
Sources within 5 arc-seconds of the planet host stars detailed in the NASA Exoplanet Archive were selected. In cases where multiple sources were found within the 5 arc-second cone around the coordinates of the planet host, the source with the lowest angular distance to the coordinates was selected.
 

\cite{KeplerPaper} have already presented updated planetary parameters for those systems discovered through the course of the \textit{Kepler} and K2 programs. As a result, we removed all \textit{Kepler}, KIC, and KOI targets from our sample, resulting in a filtered dataset containing 949 exoplanets.
The data were then used to separately perform habitable zone evaluations, radii determinations, and density determinations.
\newline \indent To calculate revised planetary radii, the dataset was first filtered to only include those systems for which the planet in question has been observed to transit its host star.
Systems with no previously reported planetary radius were also removed. 
Updated planetary radii were calculated for this revised dataset, which consisted of 320 planets.
\newline \indent 
We then proceeded to use these revised radii to determine the bulk densities of those planets for which a radius and mass was available. Of our sample of revised radii, no mass measurements were available for a total of nine planets, and so no density calculation was possible for those worlds.
\newline \indent Finally, we calculated the incident flux that would be experienced by the planets in our original dataset. We excluded those systems for which no information was available on the luminosity of the host star, or the semi-major axis of the planet in question. As a result, 259 planets were excluded from this phase of our study. This left 690 planets for which an incident flux could be calculated. \edit1{Similar to the analysis done by \cite{Kane_Habitability},} those fluxes were then compared to the boundaries for the optimistic habitable zone, as described in \citet{Kopparapu2013,Kopparapu2014}. In those works, \edit1{They provide} a mathematical description of the limits of the habitable zone for any planet, as shown in Equation \ref{eq:1}.
\begin{equation} \label{eq:1}
S_{eff} = S_{eff\odot} + aT_* + bT_*^2 + cT_*^3 + dT_*^4
\end{equation}
where $S_{eff}$ is the incident flux on the planet, $S_{eff\odot}$ is a constant corresponding to the stellar flux at the recent Venus, runaway greenhouse, maximum greenhouse, or early Mars habitable zone limit, $a$, $b$, $c$, and $d$ are polynomial coefficients determined in that work for each habitable zone limit, and $T_*$ is the host star's effective temperature.

\subsection{Revised Planet Radius}
\indent To calculate our revised planetary radii, we exploited the fact that the observed depth of a transit event
does not change with an updated distance. With a previously known stellar and planetary radius, a revised planetary radius is easily calculated. The transit depth of an event varies with the ratio of the area of the planetary disk to the area of the stellar disk as in the following equation:
\begin{equation} \label{eq:2}
\Delta F = \left(\frac{R_p}{R_*}\right)^2
\end{equation}
where $F$ is the fraction of light obscured by the transiting planet, $R_p$ is the radius of the planet, and $R_*$ is the radius of the star.
Since the the transit depth is independent of the measured distance to the star, then the ratio of the planet's radius to that of its host star will be the same regardless of that star's distance from the Earth. It is therefore fair to say that:
\begin{equation} \label{eq:3}
\frac{R_p'}{R_*'}=\frac{R_p}{R_*}
\end{equation}
Where $R_p'$ and $R_*'$ are the newly determined radii for the planet and star, respectively.
This was then solved for $R_p'$.
\begin{equation} \label{eq:4}
R_p'= \left(\frac{R_*'}{R_*}\right)R_p 
\end{equation}
which was then be applied to the selected systems to calculate a revised planetary radius using the updated stellar radii resulting from the improved distance calculations offered for those stars by {\it Gaia} DR2. \edit1{A similar methodology to calculate updated planetary radius was applied by \cite{Kane_Habitability} to the TRAPPIST-1, Kepler-186, and LHS 1140 systems.}

\subsection{Planetary Densities}
With updated planetary radii calculated using data from {\it Gaia} DR2, it is possible to determine updated bulk densities for a subset of the planets detailed in the \textit{NASA} Exoplanet Archive. Such calculations could only be performed for those planets for which a mass determination was available in the literature. Of our sample, nine systems had to be excluded on the basis of there being no available mass determination, which left 311 
planets for which a density determination could be made. Planetary density was calculated using the following equation:

\begin{equation} \label{eq:9}
\rho_p=\frac{3M_p}{4\pi R_p^3}~g~cm^{-3}
\end{equation}
\deleted{where 1.33 represents the density of Jupiter in units of $g\:cm^{-3}$ and $M_p$ and $R_p$ are in units of Jupiter mass and Jupiter radii, respectively.}

\subsection{Habitable Zone}
Recent years have seen significant interest in the frequency with which planets are found that orbit their hosts within the ''habitable zone'', where stellar insolation levels are such that one might reasonably expect an Earth-like planet to be capable of hosting liquid water on its surface. Exoplanet discovery papers regularly assess whether newly discovered planets lie within the habitable zone around their host stars. With the updated stellar parameters available from {\it Gaia} DR 2, it seems prudent to reassess these claims, and to examine the known catalogue of exoplanets to see which can still be considered to lie within the habitable zones of their stars.

\indent In order to evaluate whether a 
given planet moves on an orbit within the habitable zone around its host star, we
use the habitable zone boundaries for a $1M_\oplus$ planet provided by \cite{Kopparapu2013, Kopparapu2014}. The Recent Venus and Early Mars limits were used to identify the optimistic habitable zone while the Runaway Greenhouse and Maximum Greenhouse limits were used to identify the conservative habitable zone.
Our habitable zone evaluations were governed by a planet's incident flux and the effective temperature of its host star. Each planet's incident flux was calculated using the host star's luminosity ($L_*$) and the semi-major axis of the planet's orbit, $a$. The luminosity of the host star can be expressed as

\begin{equation} \label{eq:5}
L_*=4\pi R^2 F
\end{equation}

where F is the incident flux on the planet.
Solving for flux yields
\begin{equation} \label{eq:6}
F=\frac{L_*^2}{4\pi a^2}
\end{equation}
where $R$ is replaced by $a$ and is the semi-major axis of the planet's orbit. When the luminosity is expressed in units of the Solar luminosity, and distances are measured in au, Equation \ref{eq:6} becomes
\begin{equation} \label{eq:7}
\nicefrac{F}{F_\oplus}=\frac{\nicefrac{L_*}{L_\odot}}{\left(\nicefrac{a}{au}\right)^2}
\end{equation}
where $F_\oplus$ is the incident flux on Earth, $L_\odot$ is Solar Luminosity, and au is the distance from the Earth to the Sun, or approximately $1.496\text{x}10^8$ km. Equation \ref{eq:7} allows us to express incident flux in units of \fearth, the incident flux at the Earth (also known as the Solar constant, with a value of approximately 1360 Wm$^2$).

\indent Equation \ref{eq:7} relies on an accurate measurement of the semi-major axis of the planet's orbit. The observational results presented in {\it Gaia} DR 2 typically result in relatively small changes in the distances to the stars observed, of order just a few percent. These revised distances in turn mean that the calculated luminosities of those stars must be updated - if a star is found to be farther from Earth than previously thought, the it must also be somewhat more luminous than previously calculated. That increased luminosity will clearly impact upon the incident flux we would determine for that star's planets, and therefore alter the calculated location of the habitable zone. 

It might be natural to consider that an increased luminosity for a given star would also mean an increased calculated mass, and therefore a larger orbital semi-major axis for a planet with a given orbital period. However, we note that even relatively large changes in luminosity typically infer only a very small change in a star's mass, since the luminosity varies roughly in proportion to the mass raised to the power four, as shown in Equation \ref{eq:8}. For this reason, we assume that the calculated semi-major axis for the planets in our sample will not change as a result of the new {\it Gaia} DR 2 data, and therefore opt to calculate new insolation values for the planets in our sample using their published orbital semi-major axes. 

\begin{equation} \label{eq:8}
L_* \propto M^4
\end{equation}

It should be noted here that the determination of updated masses for the stars in question is beyond the scope of this work, as such stellar masses are best derived from individual spectra \citep{Kane_Habitability,Star_Mass,Kepler_Mass,Astroseismology_Mass}, rather than from their observed distance from Earth.

\section{Discussion} \label{results}
\subsection{Classification of Selected Stars}

The majority of the stars in our sample were of spectral types F, G, and K.
This is not unexpected, since radial velocity search programs have historically focused on such ``Sun-like'' stars, since they present an abundance of narrow spectral absorption lines that facilitate precision radial velocity measurements \citep{butler96}.  Earlier-type main sequence stars typically have too few spectral features, and those that are present are rotationally broadened and are of limited use for planet search.
 
\begin{figure}[ht!]
\centering
\resizebox{\hsize}{!}{\includegraphics[width=2.28in]{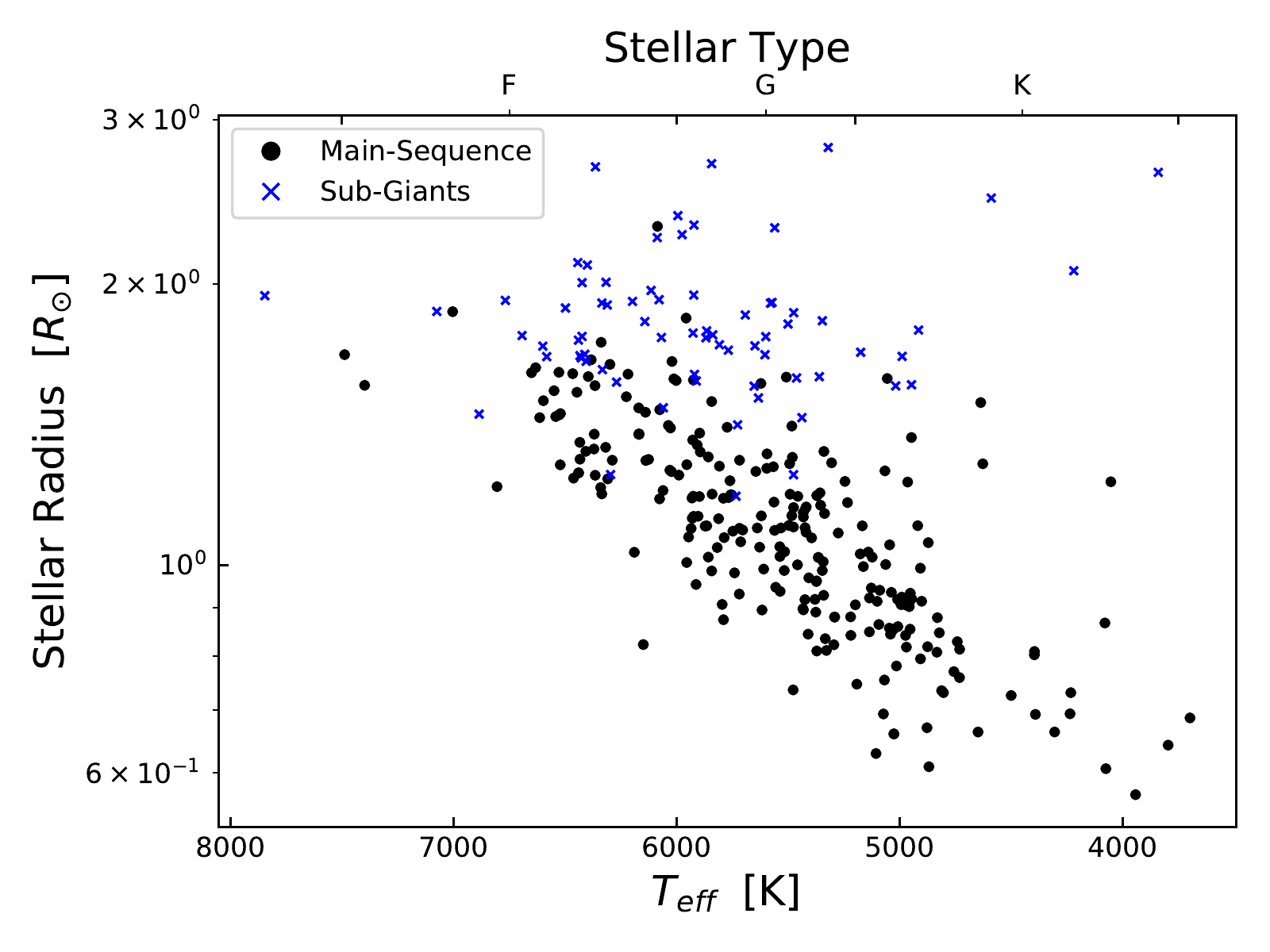}}
\caption{Stellar Classifications of 203 selected exoplanet host stars. Black dots represent main sequence stars and blue dots represent sub-giants. Only host stars with a \textit{Gaia} reported radius and temperature were plotted. Stellar radii fall between 0.6 and 3$R_\odot$. A majority of these host stars are F, G, and K type main sequence stars. Stars toward larger radii tend to span into the sub-giant branch of the HR diagram.}
\label{fig:HR}
\end{figure}

	We found that our sample of exoplanet host stars consists of 80 \% main-sequence and 20 \% sub-giants. 
This division between main-sequence and sub-giant planet hosts is illustrated in Figure \ref{fig:HR}, which shows the classification of a subset of 203 of our target stars. The division between sub-giants and main-sequence stars is clearly visible.

\begin{figure}[ht!] 
\centering
\resizebox{\hsize}{!}{\includegraphics{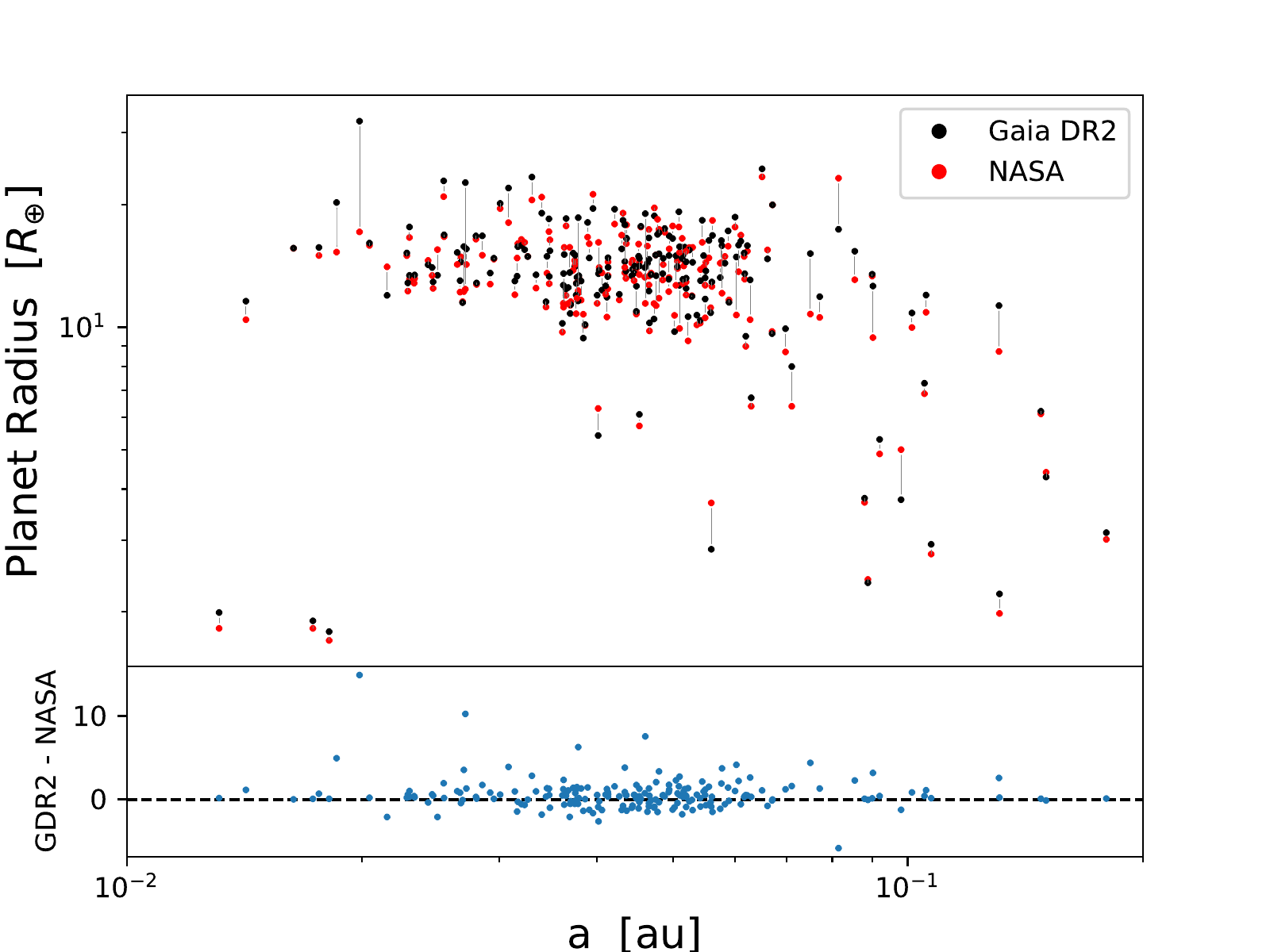}}
\caption{Planetary Radius versus Semi-Major axis. Black dots represent data presented in this work while red data represents \textit{NASA Exoplanet Archive} reported values. Grey lines represent tracks along which radius changed due to \textit{Gaia} updated luminosities. Tracks show no change in semi-major axis due to our assumption that changes in semi-major axis were negligible. For plotting purposes, the data were filtered to exclude all systems with no reported semi-major axis. The bottom panel shows the difference between revised planet radii and \textit{NASA} reported radii.}
\label{fig:au}
\end{figure}

\begin{figure}[ht!]
\centering
\resizebox{\hsize}{!}{\includegraphics[width=2.28in]{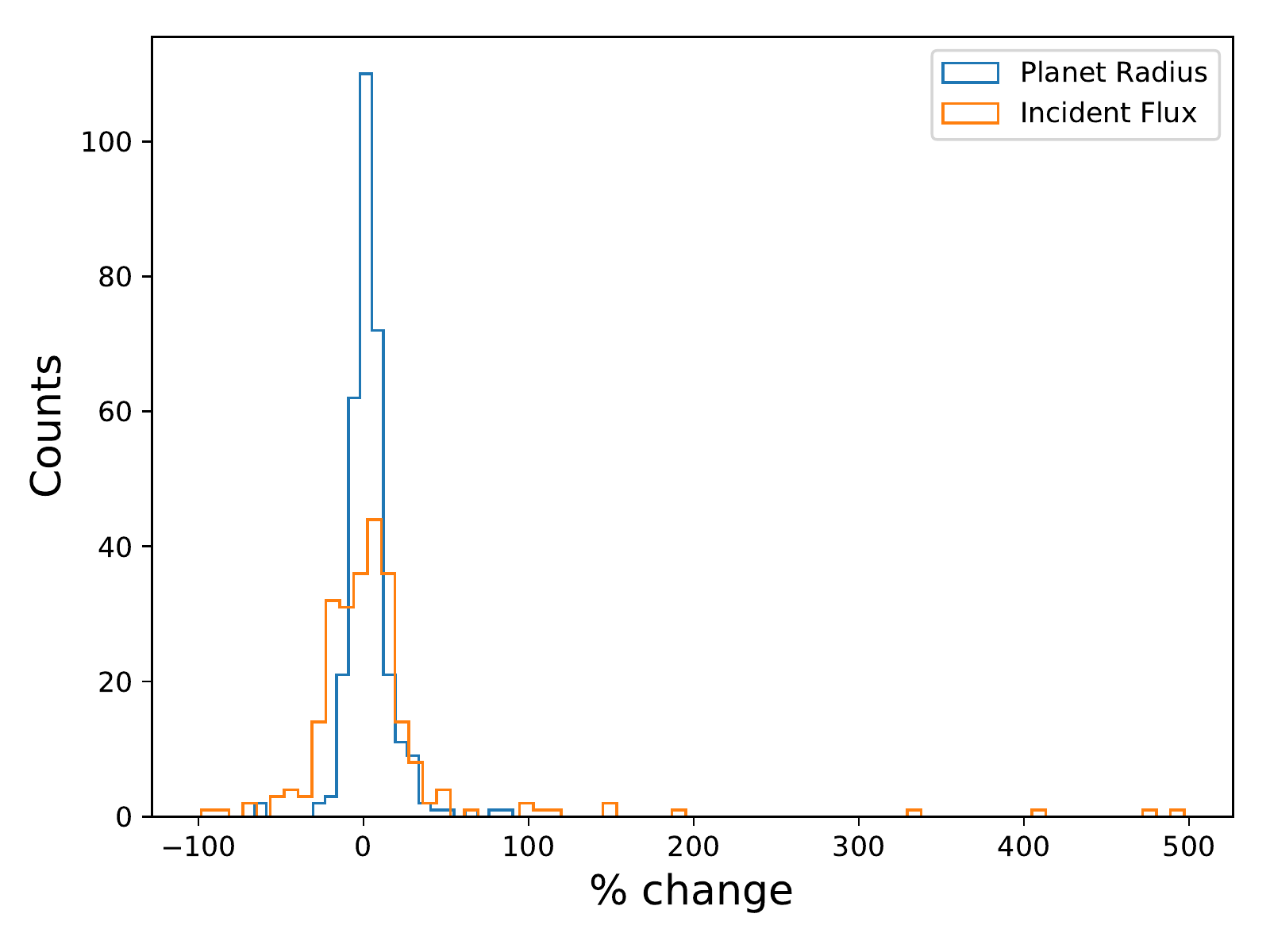}}
\caption{Histograms of the percent change in planetary radius and incident flux received by the planet. The blue histogram, which represents the percent change in planetary radius, contains 320 of our revised planet radii. The orange histogram, which represents the percent change in incident flux received by the planet, contains 690 of our revised planet incident fluxes. \edit1{We calculated a median percentage change in planetary radius and incident flux of +2.56 \% and +1.05 \%, respectively.}}
\label{fig:hist_r}
\end{figure}

\subsection{Revised Planet Radii} \label{Planets}

\begin{figure}[ht!]
\centering
\resizebox{\hsize}{!}{\includegraphics[width=2.28in]{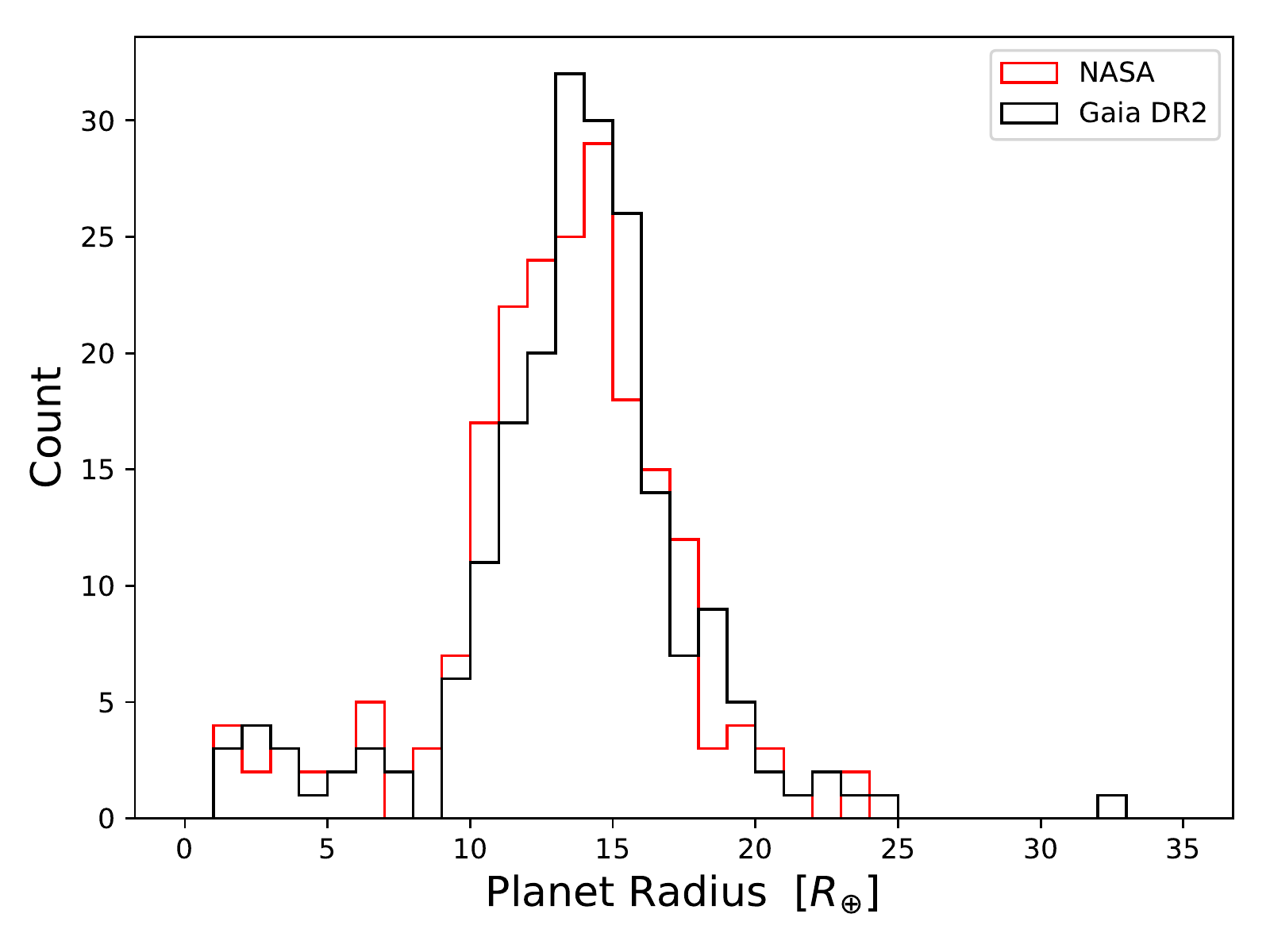}}
\caption{Distributions of planetary radii from this work (black) compared to those reported by the \textit{NASA Exoplanet Archive} (red). We find that the peak in the distribution of planetary radii, at around $14R_\oplus$ is slightly narrower and more pronounced than that seen using the old NASA Exoplanet Archive data. A K-S test of the two distributions yields a p-value of 0.0850. This indicates that the statistical distributions are likely not dissimilar.}
\label{fig:hist}
\end{figure}

\begin{figure}[ht!]
\centering
\resizebox{\hsize}{!}{\includegraphics[width=2.28in]{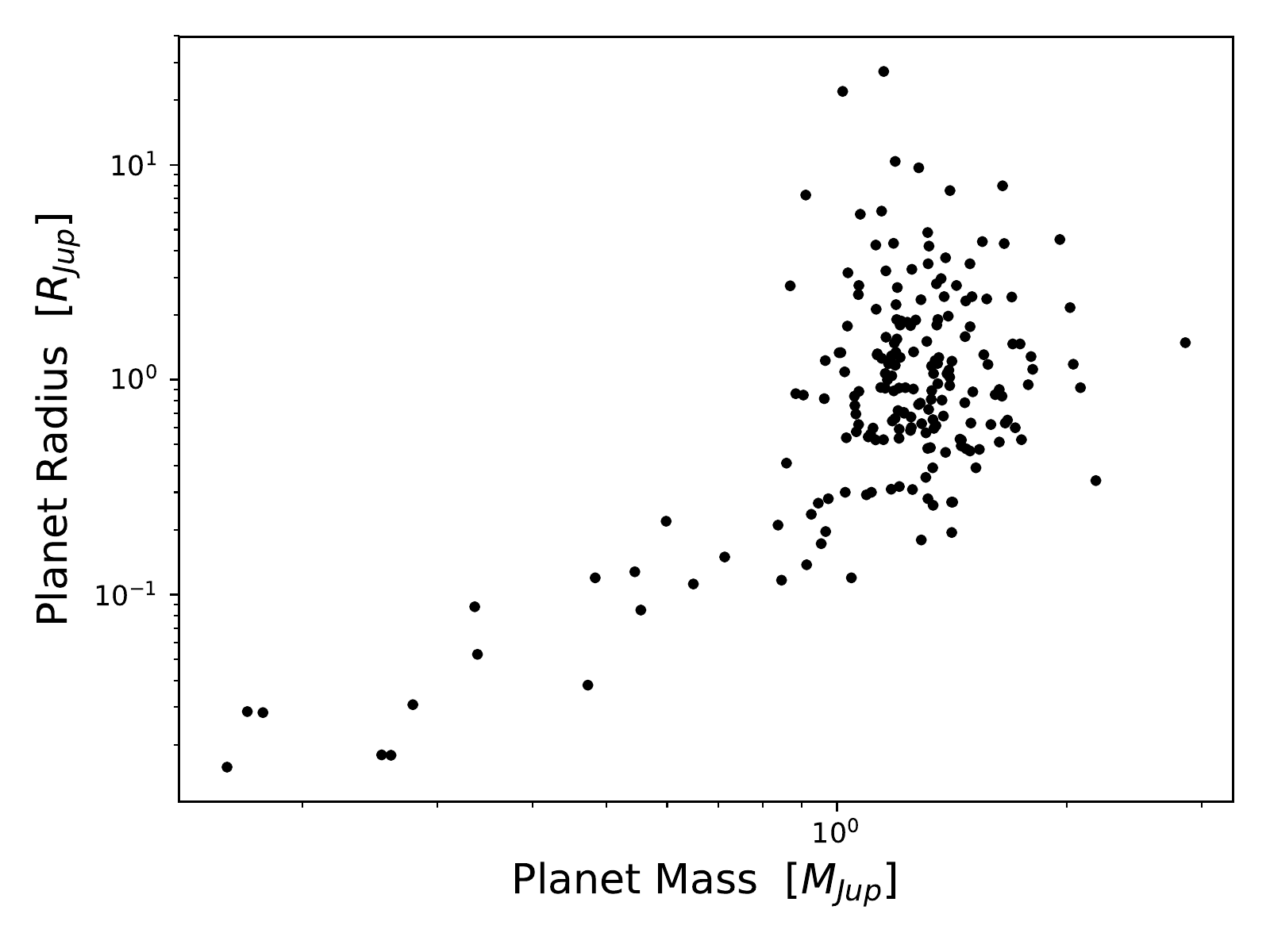}}
\caption{Planet Mass vs Revised Planet Radius of 320 planets. A large population of super Jupiters can be seen around $1M_{Jup}$ and $1R_{Jup}$. Masses are taken from the \textit{NASA} Exoplanet Archive. The clear prevalence of high mass and high radius planets is a reflection of our selection bias towards ground based observations, which are more adept at detecting Jupiter-mass planets.}
\label{fig:massVradius}
\end{figure}

\begin{figure}[ht!] 
\centering
\resizebox{\hsize}{!}{\includegraphics[width=2.28in]{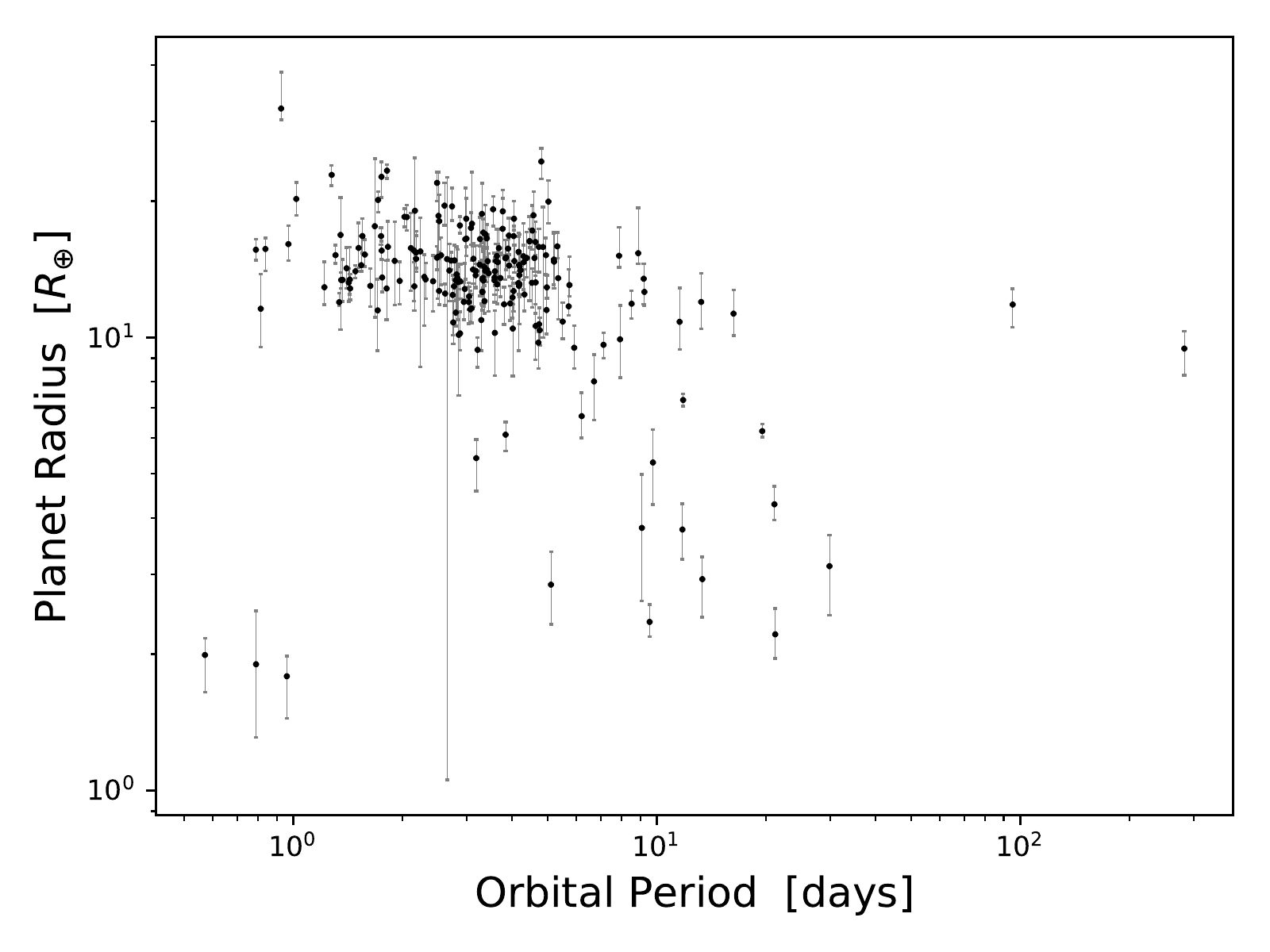}}
 \caption{Planet Radius versus Orbital Period of 320 planets. Revised planet radii are presented against orbital period to emphasize the prevalence of hot Jupiters with radii around $10R_\oplus$ and short orbital periods ranging from 1 to 10d.}
\label{fig:pd}
\end{figure}

Using \textit{Gaia} reported stellar radii we were able to calculate revised planetary radii using Equation \ref{eq:4}. The host star parameters used in this work can be found in the Appendix in Table \ref{tab:stars}. A full table of derived planetary parameters can be found in the Appendix in Table \ref{tab:planets}. We calculated a +3.76 \% average change in radius across 320 confirmed planets and a median percentage change in radius of +2.56 \%. These statistics are reflected in the bottom panel of Figure \ref{fig:au} as well as in Figure \ref{fig:hist_r}. We chose to report both the average and median percent change in planet radius due to the possibility of outliers skewing the measurement of the average percent change in radius. 
The main panel of Figure \ref{fig:au} shows a dense population of planets between 10 and $20R_\oplus$ and between 0.01 and 0.1$au$. The high density of planets around 14$R_\oplus$ is also reflected in Figure \ref{fig:hist} and is a result of our selection biased towards planets discovered using ground based observations. The average revised planetary radius was $13.61R_\oplus$ with a standard deviation of $4.21R_\oplus$. 
This population of large planets, also seen in Figure \ref{fig:massVradius} and Figure \ref{fig:pd}, are the hot and warm Jupiters - the giant planets on short period orbits to which both radial velocity and transit observations are particularly sensitive \citep{HotJupiters}.

\subsection{Revised Densities} 
\label{sec: dens}

Having revised the radii for our sample of 320 planets, we proceeded to determine the densities of those planets within that sample that had available mass determinations. Our results yielded, on average, slightly lower densities for the overall population, with the median density across our sample dropping by 6.46\%.

	Figure 7 displays planet density plotted against semi-major axis. Both the old NASA Exoplanet Archive values and revised Gaia values are plotted together to show how the distribution changed with updated parameters. There is no clear trend between semi-major axis and density.

It can be seen that most planets underwent an decrease in density, evident by the median change of -6.46\%. The average change in density was an increase of 8.64\%, and was strongly affected by planets which underwent a large decrease in radius since density scales by $R^{-3}$. One such planet, HATS-12 b, increased in density by nearly 2000\% (Table \ref{tab:planets}), and is discussed in further detail in section 3.3.2.

In Figure 8, there is a a clear increase in planet density as mass increases beyond 0.5 $M_{Jup}$. This trend agrees with the idea that above a certain mass, the gas accreted by planets compresses. The radii vary little as mass increases, which is a consequence of core electron degeneracy; hence the density increases. There is a sign of the trend of decreasing density before 0.5 $M_{Jup}$, conforming with the notion that planets will first increase in radius (and decrease in density) when they accrete gas, before gaining sufficient mass that the gas will become compressed.

\begin{figure}[ht!] 
\centering
\resizebox{\hsize}{!}{\includegraphics[width=2.28in]{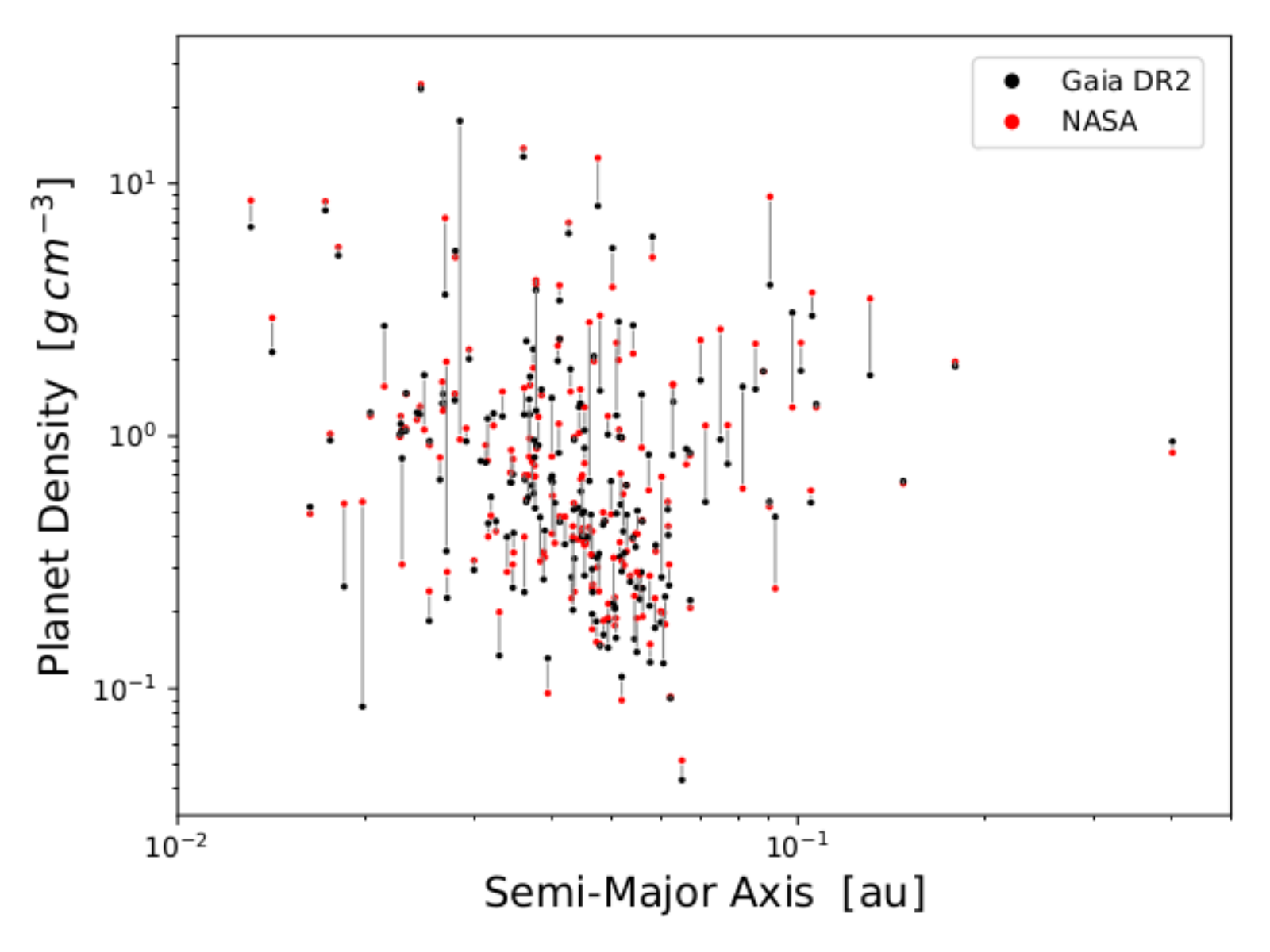}}
 \caption{Planet density versus semi-major axis. Grey lines represent tracks along which radius changed due to \textit{Gaia} updated densities.}
\label{fig:da}
\end{figure}

\begin{figure}[ht!] 
\centering
\resizebox{\hsize}{!}{\includegraphics[width=2.28in]{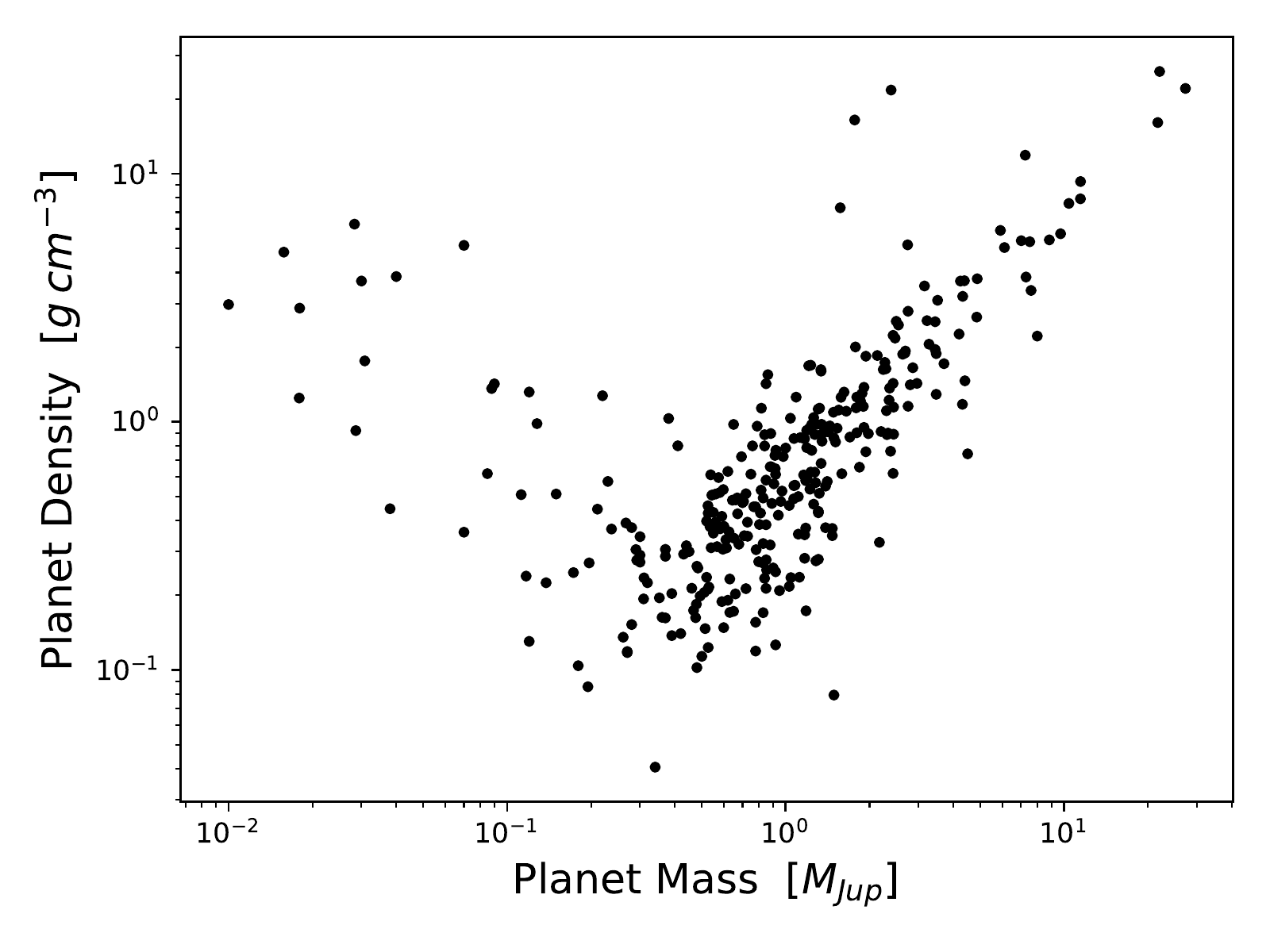}}
 \caption{Planet density versus planet mass. There is an evident trend for planets with masses $>0.5~M_{Jup}$, where density increases with increasing mass.}
\label{fig:dm}
\end{figure}

\subsubsection{WASP-103 b} 
\label{sec: wasp103}
\indent WASP-103 b had the largest radius change of any planet in our sample, increasing in radius by 87\% from the value determined by \cite{wasp103tide} to 2.86$^{+0.16}_{-0.58}$$R_{Jup}$, making it one of the largest known exoplanets. WASP-103 b is a short period exoplanet orbiting extremely close to its host star, and has been the topic of extensive transit and occultation studies \citep[e.g.][]{wasp103tide,wasp103contam,wasp103atmo}. Its orbit has been found to be on the edge of the Roche limit of WASP-103, suggesting that it may be losing mass \citep{wasp103}. 

WASP-103 b also had the second-lowest density in our sample of 0.079$^{+0.014}_{-0.049}$ $g$ $cm^{-3}$, a factor of 6 lower than the previous value of 0.55$^{+0.061}_{-0.07}$ $g$ $cm^{-3}$ due to the revised radius \citep[cf.][]{wasp103tide}. The increased radius and decreased density would appear to support the earlier hypothesis that the planet is being tidally disrupted by its host star and losing mass. However, we note that the change in distance reported for WASP-103 that resulted in these changed parameters was sufficiently large to suggest that the star itself may need to be re-examined, and that further observations are needed to confirm its parameters (Table \ref{tab:table}).

\subsubsection{High-Density Exoplanets} 
\label{sec: kelt1}
\indent In previous literature, CoRoT-3 b had the highest density of any exoplanet at 26.4$\pm5.6$ $g$ $cm^{-3}$ \citep{corot3mass}. In this work, we obtain an updated density for CoRoT-3 b of
16.1$\pm 3.98$ $g$ $cm^{-3}$. As a result, the mantle for the planet with the highest density in our sample goes to
KELT-1 b, with a new density of 22.1$^{+5.62}_{-9.16}$ $g$ $cm^{-3}$, a value slightly reduced from that published previously, of
$24.7^{+1.4}_{-1.9}$ $g$ $cm^{-3}$ \citep{kelt1mass}.\footnote{HN Peg b had a larger revised density of 25.8 $g$ $cm^{-3}$, but was excluded due to its large error of $\pm$17.9 $g$ $cm^{-3}$} Whilst definitive answers on the true density of WASP-103 b might have to wait until more observations are obtained to confirm the stellar properties, we note that that KELT-1 exhibited only a small change in luminosity as a result of its updated {\it Gaia} parameters, and so we consider it unlikely
that an updated mass will be significantly different to that published previously, see Table \ref{tab:table}. As such, we consider the density calculated for this planet robust.

Additionally, our work reveals one planet which has a revised density between KELT-1 b and CoRoT-3 b. HATS-12 b has a mass and radius markedly lower than KELT-1 b or CoRoT-3 b, either of which is massive enough to be considered brown dwarf \citep{browndwarf}. However, HATS-12 b has a mass of 2.39$\pm{0.087} M_{Jup}$, opposed to masses $>20 M_{Jup}$. It is important to note that HATS-12s Gaia luminosity decreased to 12\% of its previous value, so it is likely that the mass of HATS-12, and mass and density of HATS-12 b, also decreased (Table \ref{tab:table}). This fact again illustrates the importance of accurate determinations of stellar parameters for obtaining exoplanet characteristics.

\begin{deluxetable*}{cccccccc}
\tablecaption{The parameters of the five highest density exoplanets in our sample, based on data from {\it Gaia} DR 2. Here, $L_*$ is the luminosity of the host star, in units of the Solar luminosity, and $a$ is the semi-major axis of the planet's orbit, measured in astronomical units.} 

\tablehead{
\colhead{Planet Name}
& \colhead{Density\tablenotemark{a}} \vspace{-0.05cm}
& \colhead{Mass\tablenotemark{b}} \vspace{-0.05cm}
& \colhead{Radius\tablenotemark{c}} \vspace{-0.05cm}
& \colhead{Old $L_*$\tablenotemark{b}} \vspace{-0.05cm}
& \colhead{Updated $L_*$\tablenotemark{c}} \vspace{-0.05cm}
& \colhead{$a$\tablenotemark{b}} \vspace{-0.05cm} \\
\colhead{}
& \colhead{($g$ $cm^{-3}$)}
& \colhead{($M_{Jup}$)}
& \colhead{($R_{Jup}$)}
& \colhead{($L_\odot$)}
& \colhead{($L_\odot$)}
& \colhead{($au$)}
}
\startdata
KELT-1 b & 22.1$^{+5.62}_{-9.16}$ & 27.23$^{+0.50}_{-0.48}$ & 1.15$^{+0.10}_{-0.16}$ & 3.48$\pm0.22$ & 3.11$\pm0.05$ & 0.02466$\pm0.00016$ \\
HATS-12 b & 21.7$^{+7.80}_{-14.4}$ & 2.39$\pm0.087$ & 0.514$^{+0.060}_{-0.114}$ & 7.29$\pm1.50$ & 0.87$\pm0.11$ & 0.04795$\pm0.00077$\tablenotemark{d} \\
CoRoT-3 b & 16.1$^{+3.98}_{-3.97}$ & 21.66$\pm1$ & 1.19$\pm0.10$ & N/A & 2.46$\pm0.21$ & 0.05783$\pm0.00085$\tablenotemark{d} \\
WASP-103 b & 0.079$^{+0.014}_{-0.049}$ & 1.49$\pm0.088$ & 2.86$^{+0.16}_{-0.58}$ & 2.59$^{+0.0.39}_{-0.032}$ & 7.61$\pm1.34$ & 0.01985$\pm0.00021$ \\
\enddata
\tablenotetext{a}{Gaia DR2 revised radius and NASA Exoplanet Archive mass}
\tablenotetext{b}{NASA Exoplanet Archive}
\tablenotetext{c}{Gaia DR2 revised}
\tablenotetext{d}{NASA Composite Exoplanet Archive}
\tablecomments{Exoplanet masses are likely to change with further observations better constraining the host-star masses.}
\label{tab:table}
\end{deluxetable*}

\begin{figure*}[t]
\includegraphics[width=7.0in]{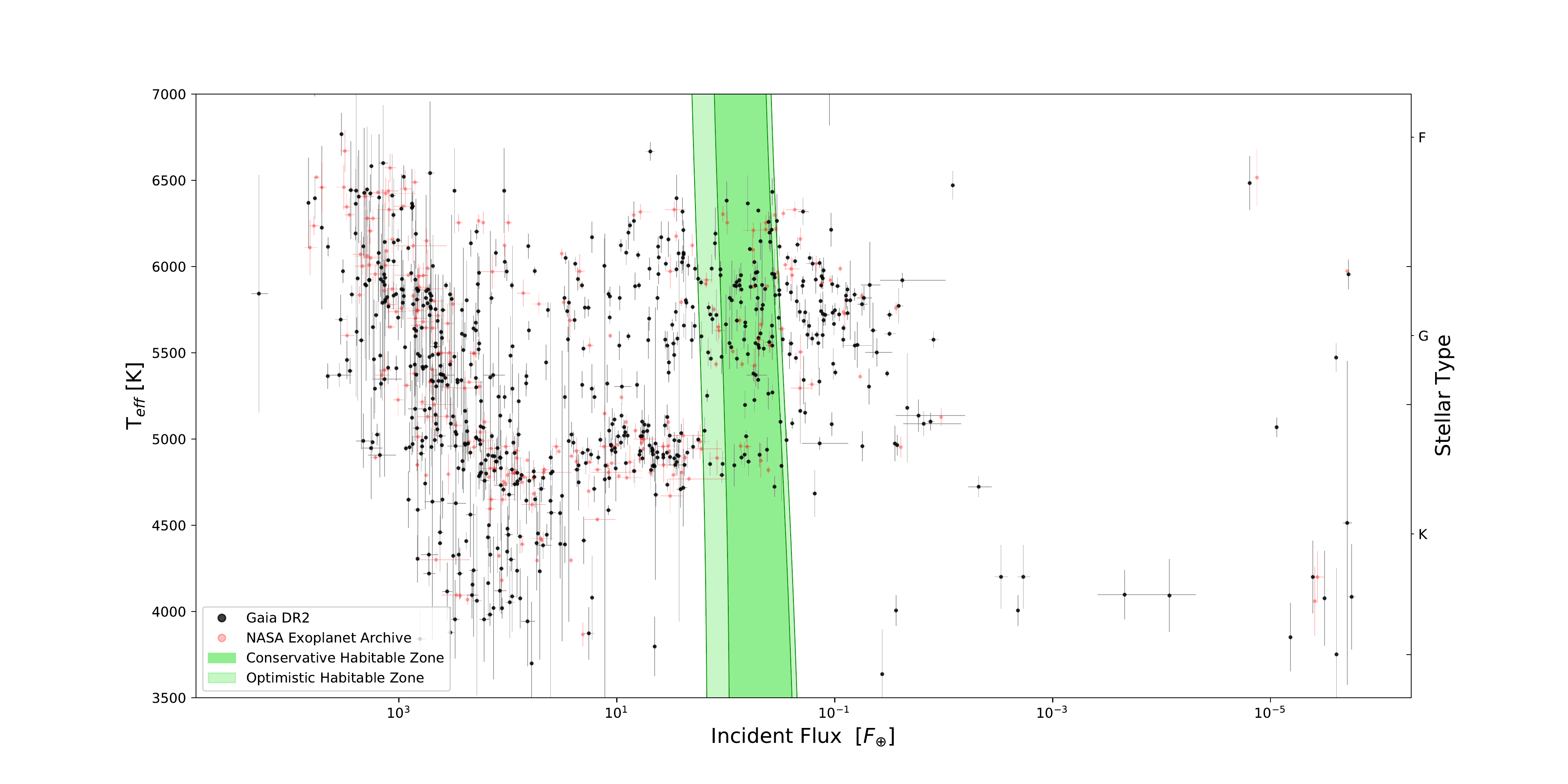}
\caption{$T_{eff}$ versus Incident Flux for 690 planets. Our data (black) is compared with previous incident fluxes (red) computed using \textit{NASA Exoplanet Archive} semi-major axes and luminosities. We plotted our error bars in black and the error bars from previously reported data in red. The wider, light green bar represents the optimistic habitable zone for an Earth mass planet. The thinner, darker green bar represents the conservative habitable zone for an Earth mass planet. Both habitable zones were identified by \cite{Kopparapu2013,Kopparapu2014}.}
\label{fig:HZ}
\end{figure*}

\begin{figure*}[t]
\includegraphics[width=7.0in]{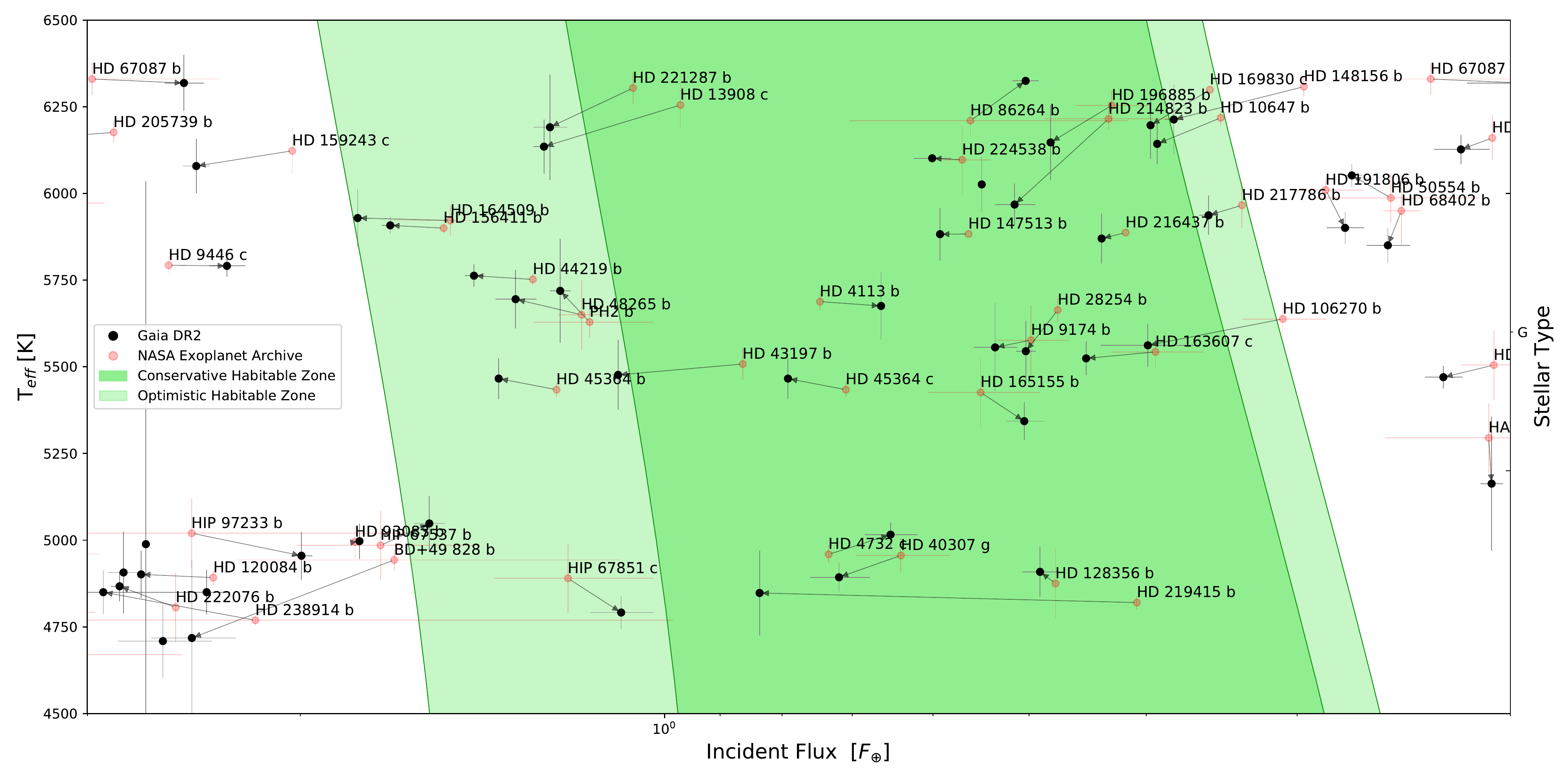}
\caption{Same as Figure \ref{fig:HZ} but with the habitable zone expanded. Points for which no previous incident flux could be calculated were removed. Planets were labeled for easy identification. A grey arrow shows a clear change in incident flux from the previous data to our revised data.}
\label{fig:HZ_zoom}
\end{figure*}

\subsection{Habitable Zones} 
\label{sec: HZ}
In order to determine whether the planets in our sample might lie within the habitable zone around their host stars, we
calculated the incident flux falling upon our sample of 690 planets. The incident flux quantifies the amount of radiant energy received by the planet from the host star, and provides some information on one of the many variables that could contribute to a given planet being considered habitable \citep[e.g.][]{Kane_Habitability,HabReview} - namely, the 'habitable zone'. In essence, for a rocky planet, or a massive satellite orbiting a giant planet, the habitable zone estimates the region around the star for which that planet could host liquid water on its surface. Since the incident flux determines the temperature of the planet, it is a good first proxy for the likelihood that a given planet could host liquid water. However, the appropriate incident flux alone does not guarantee that a given planet is habitable. A detailed discussion of the calculation of the habitable zone for stars of different spectral class, and planets of different mass, can be found in \cite{Kopparapu2014}.

Here, we adopt the 1$M_\oplus$ habitable zone limits provided by \cite{Kopparapu2013,Kopparapu2014} in order to assess which of the planets we have studied could be considered to fall within the habitable zone around their host star. Figure \ref{fig:HZ} shows all 690 confirmed planets with the optimistic habitable zone plotted in light green and the conservative habitable zone plotted in darker green. Most planets seem to reside at incident fluxes several orders of magnitude higher than those needed for habitability. This result is not, however, a surprise - both of the predominant techniques for exoplanet detection are biased towards finding planets on short-period orbits, close to their host star - and so it is unsurprising that the majority of planets in the catalog orbit at relatively small distances from their hosts. 

Figures \ref{fig:pd} and \ref{fig:HZ} show a clear grouping of planets around 10$R_\oplus$ and $150F_\oplus$, respectively. This phenomenon was also observed by \cite{KeplerPaper} and is consistent with planet inflation theory where planet radius seems to increase with incident flux through two possible mechanisms. The first mechanism of planet inflation includes the heating of the planet's interior by direct radiative heating of the planet's outer layers. The second mechanism of planet inflation includes the slowing of radiative cooling through the planet's atmosphere by the previously mentioned increase in received incident flux on the planet's surface \citep{MoreInflation,InflationTheory,Inflation}. 

Overall, we calculated an average percentage change in incident flux received by a given planet of +8.92 \%. We calculated a median percentage change in incident flux of 1.05 \%. Figure \ref{fig:HZ_zoom} shows a clear change in incident flux for planets in the habitable zone. We found that three planets that were previously located outside the habitable zone have moved into the optimistic habitable zone as a result of the {\it Gaia} DR 2 data: HIP 67537 b, HD 148156 b, and HD 106270 b. HIP 67537 was reported to have a minimum mass of $11.1^{+0.4}_{-1.1}M_{Jup}$, putting it within the planet to brown-dwarf transition region due to being above the theoretical deuterium-burning limit. HIP 67537 b also has a semi-major axis of $4.91^{+0.14}_{-0.13}\:au$, putting it on the edge of the Brown-Dwarf Desert and making it a rare object of interest \citep{HIP67537b}. HD 148156 b was reported to have a minimum mass of $9.86^{+7.77}_{-0.58}M_\oplus$ \citep{HD148156b}, making it a mini-Neptune. HD 106270 b is orbiting an evolved sub-giant and was reported to have a minimum mass of $11.0M_{Jup}$ \citep{HD106270b}. This places HD 106270 b near the theoretical deuterium-burning limit. As is outlined in \cite{HD106270b}, planets like HD 106270 b give insight into giant planet occurrence rates around more massive host stars as compared to Sun-like host stars. Occurrence rates can then be used to constrain planetary formation theories \citep{HD106270b}.

Whilst these three planets are all markedly too massive to be considered potentially rocky (and therefore habitable in their own right), it is worth considering whether they could host satellites that would, themselves, be considered potentially habitable \citep[as previously discussed in e.g.][]{AAPSOneYear,KaneMoons,HellerMoons,HillMoons}. In the Solar system, the giant planets each have a plethora of satellites, ranging from tiny, distant, irregular companions \citep[e.g][]{Irreg1,Irreg2,Irreg3} to giant regular satellites that could host sub-surface oceans \citep[e.g.][]{Europa,Callisto,Titan}. However, the total mass of the satellite systems of those four giant planets are remarkably consistent, with the combined mass of the satellites of a given planet totally approximately one ten-thousandth of the mass of the planet itself \citep[e.g.][]{Canup10k}. 

If we assume that this relationship holds true for massive exoplanets, then one would expect a giant with mass $10 M_{Jup}$ to host a satellite system with a cumulative mass of $0.1 \%$ that of Jupiter - or, in other words, approximately one third of the mass of the Earth. As such, it is perhaps not beyond the bounds of possibility that some of the most massive planets that fall within the habitable zone of their stars could host satellites that are sufficiently large to host an appreciable atmosphere.

The challenge, of course, would be to detect such a satellite. Despite the significant observational challenges involved in searching for such exomoons, it is worth noting that such efforts are currently underway \citep[e.g.][]{Kip1,Kip2,Kip3}. Those efforts have yielded tentative fruit, with the announcement of a potential satellite orbiting Kepler-1625~b \citep{exomoo}, although that satellite's existence remains controversial at this time \citep{Controv1,Controv2}. Nevertheless, if such giant satellites do indeed exist around a subset of the planets that appear to be in the habitable zone of their host stars, such moons will represent a fascinating laboratory for our ideas on planetary habitability.

\edit1{Aside from the potential for habitable exo-moons, multi-planet Jovian systems present the possibility of exo-Earths sharing the habitable zone with Jupiter size planets \citep{Agnew:2018}. As is the case with exo-moons, detecting such a planet would be near impossible with our current instruments. Future spectrographs, such as ESPRESSO \citep{ESPRESSO1,ESPRESSO2} and CODEX \citep{CODEX}, will aim for the accuracy needed (0.01-0.1~m~s$^{-1}$) to detect such a low mass object. We would like to stress that we are not claiming that the planets in this study could be habitable. Rather, we are highlighting the implications that a planet's habitability holds in the discussion of habitable exo-moons and exo-Earths.}

In addition to the three planets (or brown dwarfs) that have moved into the habitable zone in this characterisation, it is also worth noting that BD+49 828 b, which was previously reported to be in the habitable zone, was removed due to the new data yielding a significantly increased incident flux over that calculated using earlier stellar parameters. BD+49 828 b is a long period, $1.6^{+0.4}_{-0.2}M_{Jup}$ super-Jupiter orbiting an evolved Red-Giant \citep{BD+49828b}.

\section{Conclusions}
\label{conclusion}
Since derived planetary properties are so heavily dependent on the properties of their stellar hosts, which are in turn dependent on an accurate measurement of the distance to that 
star, it is critical for our understanding of the variety of exoplanets to have accurately determined distances for planet host stars. The recently released \textit{Gaia} DR2 database provides accurate parallaxes and stellar parameters for billions of stars, and represents a great improvement over the parameterisations previously available. We have used these to obtain better distance estimates for many of the stars that are known to host exoplanets, and used this to obtain updated radii and densities for many of the aforementioned exoplanets. In addition, we have used the \textit{Gaia} data to reassess whether known exoplanets orbit within the habitable zone of their host stars.
Such improved characterisation will prove vital, in the coming years, in helping to determine which planets are the most promising targets for the detailed follow-up observations that will be necessary if we are to search for any evidence of life beyond the Solar system \citep[e.g.][]{HabReview}.
\\

In summary, our key results are as follows:
\begin{itemize}
\item The updated stellar parameters from {\it Gaia} DR 2 result in an average change in planetary radius of +3.76 \% and a median change in planetary radius of +2.56 \%, across the whole sample.
\item The calculated density of CoRoT-3 b, once considered the most dense planet detected to date, is reduced from $26.4 \pm 5.6 g cm^{-3}$ to $17.3 \pm 2.9 g cm^{-3}$.
\item We report a new densest exoplanet, KELT-1 b. Its revised density was calculated to be 23.7$\pm4.0$ $g$ $cm^{-3}$. 
\item We report a revised density for WASP-103 b. WASP-103 b showed the largest change in radius of our entire sample, with the new value some 87 \% larger than that published in previous work. As a result of this increase in radius, the density of WASP-103 b was calculated to be 0.085$\pm 0.011$ $g$ $cm^{-3}$, making it one of the lowest density exoplanets of our sample.
\item We report an average percentage change in incident flux of +8.92 \% as well as a median percentage change in incident flux of +1.05 \%.
\item We report three new planets in the habitable zone based on incident flux: HIP 67537 b, HD 148156 b, and HD 106270 b. We also report one exoplanet, BD+49 898 b, being removed from the habitable zone, on the basis of a significant increase in the calculated luminosity of its host star.
\item We observed a clear prevalence of hot Jupiters in our sample, characterized by low semi-major axis, low period, and high planetary radius. This reflects a selection bias generated by picking only stars for which we had transits. Further observational studies are needed to confine the masses of these exoplanets to then provide more accurate densities.
\end{itemize}

Our work reveals the critical importance of missions such as {\it Gaia} to our ongoing attempts to better understand the variety of planets orbiting other stars. In particular, the future identification of potential targets for the search for life beyond the Solar system will rely on precise knowledge of the stars that host those planets. As {\it Gaia} continues its work, it will yield still more precise measurements of billions of stars, which will play a vital role in exoplanetary science for years to come.

\section{Acknowledgments}
\edit1{We would like to thank the referee for helpful comments on this manuscript.} This work has made use of data from the European Space Agency (ESA) mission
{\it Gaia} (\url{https://www.cosmos.esa.int/gaia}), processed by the {\it Gaia}
Data Processing and Analysis Consortium (DPAC,
\url{https://www.cosmos.esa.int/web/gaia/dpac/consortium}). Funding for the DPAC
has been provided by national institutions, in particular the institutions
participating in the {\it Gaia} Multilateral Agreement. This research has also made use of the NASA Exoplanet Archive, which is operated by the California Institute of Technology, under contract with the National Aeronautics and Space Administration under the Exoplanet Exploration Program. This material is based upon work supported by the National Science Foundation under Grant No. 1559487 and 1559505.  Additional support was provided by the Williams College Astronomy Department and Williams College Science Center.
\software{\texttt{pandas} \citep{pandas},
		  \texttt{numpy} \citep{numpy},
          \texttt{matplotlib} \citep{matplotlib},
          \texttt{SciPy}, \citep{SciPy}}

\bibliography{references}

\appendix
Contained in this appendix are the data from this work. The
host star parameters used in this work can be found in Table \ref{tab:stars}. A
full table of derived planetary parameters can be found in Table \ref{tab:planets}.

\label{tables n' stuff}
\begin{deluxetable*}{cccc}[ht!]
\tablecaption{Host Star Properties taken from \textit{Gaia} DR2 \label{tab:stars}}
\tablehead{
\colhead{Host Name}
\vspace{-0.03cm}
& \colhead{Host Luminosity}
\vspace{-0.03cm}
& \colhead{Host Radius}
\vspace{-0.03cm}
& \colhead{Host $T_{eff}$}
\vspace{-0.03cm} \\
\colhead{}
& \colhead{($L_\odot$)} 
& \colhead{($R_\odot$)} 
& \colhead{($T$)} 
}
\startdata
11 Com & 135.955$^{+3.180}_{-3.180}$ & 17.181$^{+0.555}_{-2.051}$ & 4755. \\
11 UMi & 268.853$^{+5.004}_{-5.004}$ & 30.262$^{+1.625}_{-3.414}$ & 4249. \\
14 And & 56.515$^{+0.634}_{-0.634}$ & 11.147$^{+0.282}_{-0.485}$ & 4740. \\
16 Cyg B & 1.259$^{+0.001}_{-0.001}$ & 1.120$^{+0.032}_{-0.042}$ & 5777. \\
18 Del & 37.665$^{+0.371}_{-0.371}$ & 7.911$^{+0.110}_{-0.668}$ & 5084. \\
\enddata
\tablecomments{Host name, luminosity, radius, and $T_{eff}$ of 750 exoplanet
    host stars. Host stars with no reported Radius, Luminosity, and
    $T_{eff}$ were omitted. All parameters are taken from \textit{Gaia} DR2.
    Table \ref{tab:stars} is published in its entirety in the machine-readable
        format. A portion is shown here for guidance regarding its form and
        content.}
\end{deluxetable*}

\begin{deluxetable*}{cccccc}[ht!]
\tablecaption{Revised Properties of Exoplanets \label{tab:planets}}
\tablehead{
\colhead{Planet Name}
\vspace{-0.03cm}
& \colhead{Planet Radius}
\vspace{-0.03cm}
& \colhead{Planet Mass\tablenotemark{a}}
\vspace{-0.03cm}
& \colhead{Planet Density}
\vspace{-0.03cm}
& \colhead{Semi-Major Axis\tablenotemark{a}}
\vspace{-0.03cm}
& \colhead{Planet Incident Flux}
\vspace{-0.03cm} \\
\colhead{}
& \colhead{($R_{Jup}$)} 
& \colhead{($M_{Jup}$)} 
& \colhead{($g\:cm^{-3}$)} 
& \colhead{($au$)} 
& \colhead{($F_\oplus$)}
}
\startdata
11 Com b & N/A & N/A & N/A & 1.290$^{+0.050}_{-0.050}$ & 81.699$^{+3.699}_{-1.892}$ \\
11 UMi b & N/A & 14.740$^{+2.500}_{-2.500}$ & N/A & 1.530$^{+0.070}_{-0.070}$ & 114.850$^{+5.673}_{-2.113}$ \\
14 And b & N/A & N/A & N/A & 0.830$^{+N/A}_{-N/A}$ & 82.036$^{+N/A}_{-N/A}$ \\
16 Cyg B b & N/A & 1.780$^{+0.080}_{-0.080}$ & N/A & 1.660$^{+0.030}_{-0.030}$ & 0.457$^{+0.008}_{-N/A}$ \\
18 Del b & N/A & N/A & N/A & 2.600$^{+N/A}_{-N/A}$ & 5.572$^{+N/A}_{-N/A}$ \\
\enddata
\tablenotetext{a}{NASA Exoplanet Archive}
\tablecomments{Planet name, radius, mass, density, semi-major axis, and
    incident flux of 807 exoplanets. Planets that had no calculated Radius,
    Density, and Incident Flux were omitted. 
    Table \ref{tab:planets} is published in its entirety in the machine-readable
        format. A portion is shown here for guidance regarding its form and
        content.}
\end{deluxetable*}

\end{document}